\documentclass[twocolumn,
aps,prd,
amsmath,amssymb,nofootinbib]{revtex4}
\usepackage{bm}
\newcommand{\lsim}{\mbox{\raisebox{-.9ex}{~$\stackrel{\mbox{$<$}}{\sim}$~}}}
\newcommand{\gsim}{\mbox{\raisebox{-.9ex}{~$\stackrel{\mbox{$>$}}{\sim}$~}}}

\renewcommand\({\left(}
\renewcommand\){\right)}
\renewcommand\[{\left[}
\renewcommand\]{\right]}

\newcommand\eq[1]{Eq.~(\ref{#1})}
\newcommand\eqs[2]{Eqs.~(\ref{#1}) and (\ref{#2})}

\renewcommand\eqref[1]{(\ref{#1})}

\newcommand\pa{\partial}

\newcommand\ee{\end{equation}}
\newcommand\be{\begin{equation}}
\newcommand\eea{\end{eqnarray}}
\newcommand\bea{\begin{eqnarray}}

\newcommand\mpl{M_{\rm P}}

\def\calp{{\mathcal P}}

\newcommand\bfx{{\mathbf x}}

\newcommand\sub[1]{_{\rm #1}}

\newcommand{\fnl}{{f\sub{NL}}}

\newcommand\ns{{N_\sigma}}
\newcommand\nss{{N_{\sigma\sigma}}}
\renewcommand\ne{{N\sub e}}

\newcommand\phie{{\phi\sub e}}

\newcommand\tnl{{\tau\sub{NL}}}

\newcommand\diff{{d}}

\begin{document}

\title{Generating the curvature perturbation at the end of inflation}

\author{David H.~Lyth}

\affiliation{Physics Department,
Lancaster University, Lancaster LA1 4YB,UK}


\begin{abstract}
The  dominant contribution to the primordial curvature 
perturbation may be generated at the end of inflation.
Taking the end of inflation to be sudden,  formulas are presented
for the spectrum, spectral tilt and non-gaussianity. 
They are evaluated for a minimal extension of the original 
hybrid inflation model.
\end{abstract}

\maketitle

\paragraph{Introduction}

The primordial curvature perturbation $\zeta$ exists already 
 a few Hubble times before the observable Universe
enters  the horizon.\footnote
{Standard material concerning the early Universe and observation is reviewed for
instance in \cite{treview,book}.}
 Its time-independent 
value at that stage sets the initial condition for the subsequent evolution of
all perturbations in the scalar and adiabatic mode, which in turn
seem to the dominant (perhaps the only) cause of inhomogeneity 
in the Universe. The curvature perturbation is almost gaussian, with 
an almost scale-independent spectrum 
$\calp_\zeta = (5\times 10^{-5})^2$.  

The curvature perturbation is  supposed
to originate from the vacuum fluctuations
 during inflation of one or more light 
scalar fields, which on each scale are promoted to classical perturbations
around the time of horizon exit. Different proposals have been made regarding 
when and how  this happens. In this note I explore a new possibility;
that the significant, and possibly dominant,  
 contribution to the curvature perturbation is 
 generated at the end of inflation.

\paragraph{Generating the curvature perturbation}

For clarity I assume 
 Einstein gravity,  slow-roll inflation and 
 canonically normalized fields. None of these essential for the 
 generation of  the curvature perturbation. 
What is essential though, to make the curvature perturbation almost
scale-independent, 
 is that inflation is almost exponential 
while cosmological scales leave the horizon, corresponding to 
a Hubble parameter $H$ which varies slowly on the Hubble timescale.

 The slow-roll   inflationary trajectory will lie in a subspace
of field space,  spanned by
one or more light 
fields which  are the components of the inflaton. Any additional 
light fields have  a negligible effect on the inflationary 
trajectory.

We are going to be interested in Fourier modes of the curvature
perturbation $\zeta$ with a given wavenumber $k$ of cosmological interest.
To  evaluate them we can smooth the scalar fields on a scale of order 
$1/k$ (to be precise, on a somewhat smaller scale). This scale leaves the 
horizon at some epoch $k=aH$, where $H=\dot a/a$ is the almost-constant
inflationary Hubble parameter. 

At horizon exit the canonically-normalized fields can be defined so that
 one of them  points
along the trajectory; if the inflaton has more than one component I call
this the adiabatic component of the inflaton and denote it by $\phi$,
and   call the orthogonal component(s)   isocurvature components.
If the inflaton does have more than one component, the direction of the 
trajectory changes slowly on the Hubble timescale. 

The vacuum fluctuation of each light field is promoted to a classical 
perturbation around the time of horizon exit.
The  perturbation $\delta\phi$ of the adiabatic component  (defined
just a few Hubble times after horizon exit)
gives
a contribution to $\zeta_\phi$, which remains constant until the approach 
of horizon entry. The first proposals, working in the context of 
single-component inflation,  assumed that no further contribution to $\zeta$
is generated, making $\zeta_\phi$  the observed
quantity \cite{treview,book}.
 When  multi-component inflation
was formulated it was seen  \cite{starob85}
that the isocurvature components of the
inflaton will generate an additional component to $\zeta$, but it was still
assumed that $\zeta$ will have settled down to its observed value by  the 
end of inflation. In other words, it was assumed that the curvature 
perturbation is still generated during inflation.

Later, proposals were made whereby  the dominant component of the 
curvature perturbation (or at least a significant component) is generated
only  at some 
post-inflationary epoch. According to most of these proposals the curvature
perturbation is generated at, or on the run-up to,  some reheating transition.
A reheating transition may be defined as the 
decay of  a more or less pressureless component
of the cosmic fluid (usually taken to be dominant) into radiation
to give a practically radiation-dominated Universe.
In  the first detailed proposal  \cite{curvaton} (see also \cite{earlier})
the   curvature perturbation is
generated gradually, when   the  initially
isocurvature density perturbation of some `curvaton' grows in an unperturbed
radiation background. 
According to different proposals, the curvature perturbation may be 
generated by an inhomogeneity in the decay rate \cite{zetadecay}
or the mass \cite{zetamass} 
of the decaying particle.
Further  proposals invoke inhomogeneous preheating \cite{zetapreheating},
 inhomogeneous thermalization after reheating
\cite{zetathermal} or  the inhomogeneous production of topological defects 
\cite{zetatopological}.\footnote
{The proposals of \cite{zetadecay,zetamass,zetapreheating} were first
mentioned in \cite{hk}, without any calculation.}

In this paper I explore the third logical possibility, that
 a significant and perhaps dominant component of the curvature perturbation
is  generated  at the transition between inflation and non-inflation.

\paragraph{General  formulas}

During inflation, the  light fields $\phi_i$ correspond by definition
to those directions in field space satisfying 
$|\eta_{ij}|\ll 1$, where
\be
\eta_{ij} = \mpl^2 \frac{\partial^2V/\pa\phi_i\pa\phi_j}{V}
\,.
\ee

We are assuming slow-roll inflation, which is almost exponential corresponding
to $\dot H\ll H^2$.
The  perturbations of the light fields  can be  defined on any spacetime
slicing which is non-singular \cite{treview,book}
in the limit $\dot H/H^2\to0$, such as a spatially flat slice.
These  perturbations  are almost massless and 
non-interacting. They are supposed to vanish at the classical level
before horizon exit, corresponding to the vacuum state. A few Hubble times
after horizon exit, the vacuum fluctuation has generated  classical 
perturbations $\delta\phi_i(\bfx)$, which vary
 slowly on the Hubble timescale and are practically gaussian and uncorrelated
with each other. Each of them has the same spectrum, 
 $\calp_{\delta\phi_i}=(H_k/2\pi)^2$, where $H_k$ is the value of $H$ at 
horizon exit.

The perturbations $\delta\phi_i(\bfx)$ 
  determine $\zeta$ at horizon exit, and they also determine
any  subsequent evolution of $\zeta$. 
Such evolution (until the approach of horizon entry)
 is conveniently described by the $\delta N$ formalism
\cite{starob85,ss,lms,lr,bl2} (see also \cite{st}). 
Keeping quadratic terms \cite{lr}, the
time-dependent curvature perturbation, smoothed on the scale $k$,  is
\bea
\zeta(\bfx,t)  &=& \delta N(k,\phi_i(\bfx),\rho(t))  \nonumber\\
&=& \sum_i N_i(k,t) \delta\phi_i(\bfx) + \frac12 \sum_{ij}
N_{ij} \delta\phi_i \delta\phi_j
\label{zetaofdphi}
\,.
\eea
Here, $N(k,\phi_i,\rho)$ is the number of $e$-folds, evaluated in an
unperturbed universe, from the  epoch of horizon exit when the
fields have specified values $\phi_i$,  to an
 epoch when the energy density has a specified
value  $\rho$.
In the  second line, $N_i\equiv \partial N/\partial \phi_i$
and  $N_{ij}\equiv \partial^2 N/\partial \phi_i\pa \phi_j$. 
These derivatives are evaluated on the unperturbed trajectory, on which
the  $\phi_i$ are determined by  $k$ making $N_i$ and $N_{ij}$ functions
of just $k$ and the final time $t$.

At some stage before nucleosynthesis,  $\zeta$ settles down to a
time-independent value, which is constrained by observation.
We write down the predictions for $\zeta$ which follow from \eq{zetaofdphi},
as a function of
 $t$ even though we are interested in the regime where $\zeta$
has settled down to its final value.

Since the observed $\zeta$ is almost gaussian, one or more linear terms
must dominate \eq{zetaofdphi} giving the spectrum
\be
\calp_\zeta(k,t)  = \sum_i N^2_i(k,t)  (H_k/2\pi)^2
\label{multispec}
\,.
\ee
The  spectral tilt $n-1\equiv \pa \ln \calp_\zeta/\pa \ln k$
is  \cite{ss,treview}
\be
n-1 =  2\frac{\eta_{jm} N_j N_m }{N_n N_n} -2\epsilon -\frac2{\mpl^2 N_iN_i} 
\,,
\label{multitilt}
\,.
\ee
where   identical indices are summed over. Here $\epsilon\ll 1$
is a slow-roll parameter, defined as $\frac12 \mpl^2(V'/V)^2$ where
$V'$ is the derivative of $V$ along the inflationary trajectory.
Both $\epsilon$ and $\eta_{jm}$ are to be evaluated at horizon exit.

If the adiabatic component of the inflaton  dominates \eq{zetaofdphi},
 the non-gaussianity of $\zeta$
is  too small  to ever be observable  \cite{maldacena,sl}. Otherwise
it may be observable. The likely observables are
 the bispectrum and trispectrum, which alone are generated by the quadratic
expansion \eqref{zetaofdphi}. They are specified respectively \cite{bl2}
by quantities
$\fnl$ and $\tnl$. Taking the field perturbations to be perfectly gaussian,
which has been justified for the bispectrum \cite{sl,lz}, and ignoring
the scale-dependence of $\calp_\zeta$, the predictions
are \cite{lr,bl2}\footnote
{The general
formula for $\tnl$ follows from the special cases
in \cite{bl2} but has not been written down before. For
$\fnl$, the  scale-dependence
of the spectra is taken into account in \cite{dg}.}
\bea
-\frac35\fnl &=&
\frac12 \frac{  N_i N_{ij} N_j }{ (N_n N_n)^2 }
+ 4A \calp_\zeta \frac{ {\rm Tr\,} N^3 }{ (N_m N_m)^3 }
\label{fnl} \\
\tnl &=& 2 \frac{  N_i N_{ij} N_{jk} N_k }{(N_n N_n )^3 }
+ 16B \calp_\zeta \frac{ {\rm Tr\,} N^4 }{ ( N_m N_m)^4 }
\label{tnl}
\,,
\eea
where $A$ and $B$ are of order 1 on cosmological scales.
Present observation \cite{fnlbound}
gives roughly $|\fnl|\lsim 100$, and absent a detection the eventual
bound will be \cite{bartolrev} $|\fnl|\lsim 1$. There is at present
no bound on $\tnl$ from modern data, and no estimate of the bound that will
eventually be possible. (A crude bound from COBE data \cite{bl2}
 is $|\tnl|<10^{8}$.)

\paragraph{Dominance by a single field perturbation}

If only one  field $\sigma$ is relevant,
 \eq{zetaofdphi}
  becomes in an obvious notation
\be
\zeta(t,\bfx) = \ns \delta\sigma  + \frac12 \nss (\delta\sigma)^2
\label{simple1}
\,,
\ee
leading to
\bea
\calp_\zeta & \simeq & \ns^2 (H_k/2\pi)^2 \label{spec} \\
n-1 & \simeq &  2\eta_{\sigma\sigma} -2\epsilon -\frac2{\mpl^2 \ns^2} 
\label{specin}
\,.
\eea
Because the first term of \eq{simple1} dominates, 
\be
-\frac35\fnl = \frac12 \frac{\nss}{\ns^2}
\label{simple2}
\,,
\ee
and \cite{bl2} $\tnl = 36\fnl^2/23$.
In this case, $\fnl$ may equivalently be defined
by writing $\zeta = \zeta\sub g - \frac 35 \zeta\sub g^2$
where $\zeta\sub g$ is gaussian.\footnote
{Following \cite{maldacena,lr,bl2} I am  defining
 $\fnl$ through the curvature perturbation. An alternative definition of 
$\fnl$ works with the Bardeen potential.
The two definitions coincide in  first-order perturbation theory but differ
at second order.}

If $\sigma$ is the adiabatic component $\phi$ of the inflaton, $\delta\phi$
is just a shift back and forth along the trajectory, making $\delta N$
time-independent with
\bea
\mpl^2 N^2_\phi &=& \frac1{2\epsilon}\label{nsphi}  \\
N_{\phi\phi} &=& (2\epsilon - \eta) N^2_\phi \label{nphis}
\,,
\eea
where   $\eta\equiv \mpl^2 V''/V$ is another slow-roll parameter.
Both $\epsilon$ and $\eta$ are to be evaluated at horizon exit in these
expressions.

Putting  \eq{nsphi} into \eqs{spec}{specin} reproduces the usual expressions
\bea
\calp_\zeta & =& \frac1{2\mpl^2\epsilon} \( \frac{H_k}{2\pi} \)^2 \\
n-1 &=& 2\eta - 6\epsilon
\,.
\eea
Putting \eqs{nsphi}{nphis} into \eq{simple2} gives $|\fnl|\ll 1$. This
means \cite{sl,lr} that the non-gaussianity is too small to observe, as was
first demonstrated in a different way by Maldacena \cite{maldacena}.
In this case the tensor fraction of the density perturbation is
$r=16\epsilon$.\footnote
{I adopt the current definition of $r$ in terms of the spectra. An older
definition using the CMB quadrupole corresponds to $r=12.4 \epsilon$.}
The tensor may  eventually be observable \cite{rlimit} if $r\gsim 10^{-4}$.

If $\sigma$ is some other field, its contribution compared with that of
$\phi$ will be
\be
\( \frac{\zeta_\sigma}{\zeta_\phi} \)^2 \sim \frac{\calp_{\zeta_\sigma}}
{\calp_{\zeta_\phi}}
 = \frac{N_\sigma^2}{N_\phi^2}
\gg 1
\,.
\ee
Then the middle term of \eq{specin} becomes negligible giving
\be
n-1 =  2\eta_{\sigma\sigma} - 2\epsilon
\label{ncurv}
\,.
\ee
 Since the tensor perturbation
depends only on $H$ the tensor fraction $r$ is reduced;
\be
r = 16\epsilon  \frac{\calp_{\zeta_\phi}}{\calp_{\zeta_\sigma}}
  \ll 16\epsilon
\,.
\ee
It will never be observable if $\calp_{\zeta_\phi}/\calp_{\zeta_\sigma}$
 is bigger than about $10^5$.
On the other hand, 
non-gaussianity given by \eq{simple2} could be observable.

\paragraph{Generating some curvature perturbation at the end of inflation}

To see how a contribution to the curvature perturbation may be generated at the
end of inflation,  consider first single-component inflation, 
where there is a unique inflationary trajectory $V(\phi)$,
giving $\phi$ as a unique function of. 
 proper time $\tau$ up to a shift in the origin.
 All functions of
$\phi$ and its derivatives will also be unique, including the energy 
density $\rho=V+\frac12\diff^2\phi/\diff\tau^2$.

At each position 
inflation ends when $\phi$ has some value  $\phie$.
Usually it is assumed that the end of inflation is controlled entirely
by the inflaton, so that $\phie$ is independent of position. 
Then inflation ends
on a spacetime slice of uniform energy density.
The new possibility, considered here for the first time, is that 
$\phie$ depends on some field  $\sigma$, whose potential is so flat that it
has practically no effect on the inflaton trajectory (consistent with the
assumption of single-component inflation). Then $\phie(\sigma)$ 
will depend on position through the perturbation $\delta\sigma(\bfx)$.
As a result the change $N\sub{ba}$,
 from  a spacetime slice of uniform density just {\em before} the end of 
inflation, to a spacetime slice of uniform density just {\em after} the end of 
inflation will have a perturbation $\delta N\sub{ba}\equiv 
\zeta\sub e(\bfx)$.
(Indeed, $N\sub{ba}$ will bigger, the bigger is the 
delay in the end of inflation.)
This quantity $\zeta\sub e$
is the contribution to the curvature perturbation generated by the 
end of inflation, which might dominate the inflaton contribution
$\zeta_\phi$.

In  the case of multi-component inflation there is  a 
family of curved inflationary trajectories in, say, a two-dimensional 
field space. In our   part of the Universe one of the trajectories
is the unperturbed trajectory,  but the perturbation
in the field  orthogonal to the trajectory at horizon exit kicks the local
evolution onto nearby trajectories, causing 
 $\zeta$ to  vary  after horizon exit. 

If $\zeta$ has settled down
to a constant value by the end of inflation, corresponding to all trajectories
becoming essentially equivalent straight lines, 
the generation of a contribution to the 
curvature perturbation at the end of inflation goes through in exactly the
way that we described for the single-component case. 
In particular, $\zeta\sub e$ vanishes unless some field other than the,
by now essentially unique, inflaton comes into play.

If instead 
 the family of trajectories is still curved as
the end of inflation is approached, the situation is quite different.
There is now no reason for inflation to end on a slice of uniform density,
and one expects that $\zeta\sub e$ will be nonzero even if the 
multi-component inflaton completely determines the end of inflation.
In the case of hybrid inflation, this is because one expects different 
components of the inflaton to have different couplings to the waterfall
field. In the case of  non-hybrid  inflation, 
 the end of inflation on a given local trajectory
will occur when at least one of the 
 parameters $\epsilon$ or $\eta$ has a value of order 1. But that value
depends  on the local trajectory which (in contrast with single-component
inflation) depends on position, and in any case 
 the slices of uniform
$\epsilon$, uniform $\eta$ and uniform $\rho$ will all be different.
Again, we see that $\zeta\sub e$ is expected to be nonzero, just because
there is a family of inflationary trajectories which are still 
inequivalent as the end of inflation approaches.

\paragraph{The sudden-end approximation}

To have a clean result, let us suppose that  slow-roll inflation suddenly
gives way to radiation domination. This  case 
may be realistic in a hybrid inflation model where the waterfall field is quite
heavy, and in any case the calculation should give a feel for the sort of
effect that might be possible. We will denote quantities evaluated
at the end of inflation by a subscript e.

As slow-roll persists right up to the end, expansion during inflation is
far more rapid than it is afterward, and $\zeta\sub e$ is practically equal
to the number of extra $e$-folds of inflation. Expanding to second order in
the field perturbations, the perturbation in $\phie(\sigma)$ is
\be
\delta\phie = \phie' \delta\sigma + \frac12 \phie'' (\delta\sigma)^2
\,.
\ee
During inflation, the 
derivatives of $N(\phi)$ are given by \eqs{nsphi}{nphis} and we 
denote them respectively by $\ne'$ and $\ne''$. Then
\bea
\zeta\sub e &=& \ne' \delta\phie + \frac12 \ne'' (\delta\phie)^2 \\
&=& \ne' \phie' \delta\sigma + 
\frac 12 \[ 2 \ne'' {\phie'}^2 + \ne'\phie'' \] (\delta\sigma)^2 
\label{zetasudden}
\,.
\eea

Let us assume that $\zeta\sub e$   dominates $\zeta_\phi$, corresponding to
\be
{\phie'}^2 \gg \frac {\epsilon\sub e}{\epsilon_k}
\,.
\ee
Then \eq{zetasudden} is dominated by the first term, giving 
\be
\calp_{\zeta\sub e} = \frac {  {\phie'}^2 } {2\epsilon\sub e} 
\( \frac {H_k} {2\pi} \)^2
\,.
\ee
The spectral index is given in terms of $V_{\sigma\sigma}$ by
\eq{ncurv}.
 The non-gaussianity is given by \eqref{simple2} 
(imposing $\epsilon\sub e\ll 1$ and  $|\eta\sub e|\ll 1$) as 
\be
-\frac 35 \fnl = \frac{\sqrt{2\epsilon\sub e}} 2
\frac{\phie''}{ (\phie')^2 }
\,,
\ee
which could be observable.

\paragraph{A concrete model}

Let us apply these formulas to a concrete model of inflation;
\bea
V &=& V_0 - \frac12m_\chi^2\chi^2  +\frac14 \lambda \chi^4 
+\frac12 m_\phi^2\phi^2 
+\frac12\lambda_\phi\phi^2\chi^2 \nonumber \\
 &+& \frac12\lambda_\sigma \sigma^2\chi^2 +V_\sigma(\sigma)
\,.
\eea
This is the original hybrid inflation model \cite{originalhybrid},
except for the last term two terms. One of them
 specifies the interaction of $\sigma$
with the waterfall field $\chi$, and the other gives its potential during 
inflation. 
 The interaction corresponds to an inhomogeneous mass-squared for the 
waterfall field;
\be
-m_\chi^2(\sigma) = -m_\chi^2 + \lambda_\sigma \sigma^2
\,,
\ee
which is  required to be negative.

During inflation the waterfall field $\chi$ is pinned at the origin.
leaving just $V_\sigma(\sigma)$ which we assume makes $\sigma$ light
($|\eta_{\sigma\sigma}| \ll 1$) and has  negligible slope.
Inflation ends when $\chi$ is destabilized,
corresponding to 
\be
\lambda_\phi\phie^2 + \lambda_\sigma \sigma^2 = m_\chi^2
\,.
\ee

The only effect of the extra term is to make $\phie$ a function of 
the new field $\sigma$. As in the original model, we require
$\phie\ll \mpl$,  so that inflation after the observable Universe leaves
the horizon takes place entirely in the regime $\phi\ll \mpl$. As a result
$\epsilon$ is completely negligible, which has two consequences.
First the tensor fraction is  unobservably small \cite{al}. Second
the spectral tilt, if  big enough to be observable,
 will be 
given by just the first term of the appropriate equation \eqref{multitilt},
\eqref{specin} or \eqref{ncurv}.

The derivatives of $\phie(\sigma)$ are
\bea
\phie' &=& - \frac{\lambda_\sigma \sigma}{\lambda_\phi \phi\sub e}\\
\phie'' &=& - \frac{\lambda_\sigma }{\lambda_\phi \phi\sub e}
\( 1 - \frac{\lambda_\sigma \sigma^2}{\lambda_\phi \phi^2\sub e}
\)
\,.
\eea
Putting them into \eq{zetasudden} gives $\zeta\sub e$, whose spectrum is
\be
\calp_{\zeta\sub e} = \frac1{2\epsilon\sub e}
\( \frac{\lambda_\sigma \sigma}{\lambda_\phi \phi\sub e} \)^2
\( \frac {H_k} {2\pi} \)^2
\,.
\ee
The condition for $\zeta\sub e$   to dominate $\zeta_\phi$ is\footnote
{Remember that $N\sub e$, the value of $N$ at the end of inflation, is
called simply $N$ when one discusses unperturbed inflation with 
a value $N\simeq 60$. I am assuming $V\simeq V_0$ as in the original model.}
\be
\( \frac{\lambda_\sigma \sigma}{\lambda_\phi \phi\sub e} \)^2
\gg \frac{\epsilon\sub e}{\epsilon_k} = e^{-2N_e\eta}
\label{domcon}
\,,
\ee
with $\eta = m_\phi^2\mpl^2/V_0$. If it does dominate,
 the spectral index is given
by \eq{specin} and has nothing to do with $\eta$. 
Therefore $\eta$ can have any value $\ll 1$, and for a crude order of magnitude
estimate it should even be possible to consider $\eta\sim 1$.
As a result the right hand side of \eq{domcon} can be far below 1, and so
can the left hand side.

Let us assume that  $\zeta\sub e$ dominates. Then
 the spectral tilt, if observable, is
$n-1 = 2\eta_{\sigma\sigma}$. The 
 non-gaussianity parameter, if observable,  is
\be
-\frac35 \fnl = 
\frac{ \eta}{2}
 \(  
\frac{
\lambda_\phi \phi^2\sub e
}{
\lambda_\sigma \sigma^2
}
\)
\,.
\ee

The predictions of the model involve the unperturbed field value 
$\sigma$ and its effective mass parameter $\eta_{\sigma\sigma}$.
The unperturbed value should be taken as   the mean of $\sigma(\bfx)$
within the observable Universe, as opposed to its value in some 
exponentially larger region \cite{bl2}.
This situation has been discussed
 previously for the curvaton 
\cite{curvaton,dlnr} (see also  \cite{andreiaxion} for the axion).
As  $\sigma$ is light 
it  will not be trapped at a minimum of the potential even if inflation
has lasted for
a long time before the observable Universe enters the horizon.
However in that case 
there will be a calculable probability distribution for $\sigma$,
and one is then free to assume that the observable Universe is typical
which gives the rough value of $\sigma$.
Because inflation does not `know' about $\sigma$ (until the very end) it may
be regarded as quite  likely that
 the lightness condition $|\eta_{\sigma\sigma}|\ll 1$ will 
 be very well satisfied, making the spectral tilt too small to observe.
(This is in contrast with many inflation models where $\eta$ increases 
as inflation proceeds giving a spectral tilt of order $1/N$ \cite{al}.)
If the spectral tilt is big enough to observe
 there is some prejudice towards $n>1$ because the
 peak of the probability distribution is at the minimum of $V(\sigma)$,
but $n<1$ is also quite reasonable at least if
 $\sigma$ is a pseudo-Goldstone-Nambu-boson corresponding to a periodic
potential.

\paragraph{Conclusion}

The existence of  many scalar fields is mandatory if the field theory
is supersymmetric, and is anyhow suggested by string theory even if the 
field theory is non-supersymmetric. Flat directions 
in field space, corresponding to a small quadratic term in the potential
and a negligible quartic term,  can occur  in  
supersymmetric and  conformal \cite{paul} field theories. 
As a result, it is reasonable to suppose  that the vacuum fluctuation of 
several scalar fields could be promoted at horizon exit to become a classical
perturbation, and to expect that one will
 find a variety of possible 
mechanisms for generating the curvature perturbation.

It has been known for a long time that the curvature perturbation 
changes
with time if and only if the function $P(\rho)$ is inhomogeneous. It follows
that a (more or less sharp)
 phase transition, corresponding  to a  change in this 
function, will generate some curvature perturbation if and only if it occurs
on a spacetime slice which is {\em not} one of uniform $\rho$. That is one
possible way of generating some of the curvature perturbation, or practically
all of it. The well-known example \cite{zetadecay}
is a reheating transition ($P\simeq 0$ to $P=\rho/3$),
 and I have explored another example which is the transition
at the end of inflation ($P\simeq -\rho$ to $P\leq \rho/3$). 
Judging by the hybrid inflation example, the new mechanism is very easy
to implement, and can take place for a wide range of the parameters.

After the first version of this note was released, I was made aware that
a  different model generating a contribution to the curvature perturbation 
at the end of hybrid inflation was proposed in \cite{bku}. The model 
invokes an inhomogeneous coupling instead of an inhomogeneous mass. Also,
the calculation is done by matching metric perturbations whereas I am using
the $\delta N$ formalism.

{\em Acknowledgments.}
The author is 
supported by PPARC grants PPA/G/O/2002/00469,  PPA/V/S/2003/00104,
PPA/G/O/2002/00098 and PPA/S/2002/00272 and by EU grant 
 MRTN-CT-2004-503369.


\begin{thebibliography}{}

\bibitem{treview} 
D.~H.~Lyth and A.~Riotto, Phys. Rep. {\bf 314}, 1 (1999).

\bibitem{book}
A.~R.~Liddle and D.~H.~Lyth, \emph{Cosmological Inflation and            
Large Scale Structure}, (CUP, Cambridge, 2000).

\bibitem{starob85}
 A.~A.~Starobinsky,
  JETP Lett.\  {\bf 42} (1985) 152
  [Pisma Zh.\ Eksp.\ Teor.\ Fiz.\  {\bf 42} (1985) 124].
\bibitem{curvaton} 
D.~H.~Lyth and D.~Wands,
Phys. Lett. B {\bf 524}, 5 (2002);
T.~Moroi and T.~Takahashi,
Phys. Lett. B {\bf 522}, 215 (2001)
[Erratum-ibid.\ B {\bf 539}, 303 (2002)].

\bibitem{earlier}
S.~Mollerach,
Phys. Rev. D {\bf 42}, 313 (1990);
A.~D.~Linde and V.~Mukhanov,
Phys. Rev. D {\bf 56}, 535 (1997);
K.~Enqvist and M.~S.~Sloth, Nucl. Phys. B {\bf 626}, 395 (2002).

\bibitem{zetadecay}
 G.~Dvali, A.~Gruzinov and M.~Zaldarriaga,
  Phys.\ Rev.\ D {\bf 69} (2004) 023505
 L.~Kofman,
  arXiv:astro-ph/0303614.

\bibitem{zetamass}
 G.~Dvali, A.~Gruzinov and M.~Zaldarriaga,
  Phys.\ Rev.\ D {\bf 69} (2004) 083505

\bibitem{zetapreheating}
  M.~Bastero-Gil, V.~Di Clemente and S.~F.~King,
  Phys.\ Rev.\ D {\bf 70} (2004) 023501
 E.~W.~Kolb, A.~Riotto and A.~Vallinotto,
  Phys.\ Rev.\ D {\bf 71} (2005) 043513

\bibitem{zetathermal}
 C.~W.~Bauer, M.~L.~Graesser and M.~P.~Salem,
  Phys.\ Rev.\ D {\bf 72} (2005) 023512

\bibitem{zetatopological}
T.~Matsuda,
  arXiv:hep-ph/0509063.


\bibitem{hk}
  T.~Hamazaki and H.~Kodama,
  Prog.\ Theor.\ Phys.\  {\bf 96} (1996) 1123


\bibitem{lr}
 D.~H.~Lyth and Y.~Rodriguez,
  Phys.\ Rev.\ Lett.\  {\bf 95}, 121302 (2005).

\bibitem{bl2}
  L.~Boubekeur and D.~H.~Lyth,
  arXiv:astro-ph/0504046.


\bibitem{ss} M.~Sasaki and E.~D.~Stewart,
Prog. Theor. Phys. {\bf 95}, 71 (1996).

\bibitem{lms}
 D.~H.~Lyth, K.~A.~Malik and M.~Sasaki,
  JCAP {\bf 0505} (2005) 004

\bibitem{st}
  M.~Sasaki and T.~Tanaka,
  Prog.\ Theor.\ Phys.\  {\bf 99}, 763 (1998).


\bibitem{dlnr}
 K.~Dimopoulos, D.~H.~Lyth, A.~Notari and A.~Riotto,
  JHEP {\bf 0307} (2003) 053

\bibitem{maldacena}
J.~Maldacena, 
JHEP {\bf 0305}, 013 (2003).

\bibitem{rlimit}
  M.~Amarie, C.~Hirata and U.~Seljak,
  arXiv:astro-ph/0508293.

\bibitem{sl} D.~Seery and J.~E.~Lidsey, astro-ph/0503692.

\bibitem{lz}
 D.~H.~Lyth and I.~Zaballa, JCAP {\bf 10} (2005) 005.

\bibitem{dg}
L.~E.~Allen, S.~Gupta and D.~Wands,
  arXiv:astro-ph/0509719.

\bibitem{fnlbound}
  P.~Creminelli, A.~Nicolis, L.~Senatore, M.~Tegmark and M.~Zaldarriaga,
  arXiv:astro-ph/0509029.

\bibitem{bartolrev}
 N.~Bartolo, E.~Komatsu, S.~Matarrese and A.~Riotto,
  Phys.\ Rept.\  {\bf 402} (2004) 103.

\bibitem{originalhybrid}
 A.~D.~Linde,
  Phys.\ Lett.\ B {\bf 259} (1991) 38.

\bibitem{andreiaxion}
 A.~D.~Linde,
Phys.\ Lett.\ B {\bf 201}, 437 (1988).

\bibitem{al}
 D.~H.~Lyth and L.~Alabidi,
  arXiv:astro-ph/0510441.

\bibitem{paul}
   P.~H.~Frampton and T.~Takahashi,
  Phys.\ Rev.\ D {\bf 70} (2004) 083530.

\bibitem{bku}
  F.~Bernardeau, L.~Kofman and J.~P.~Uzan,
  Phys.\ Rev.\ D {\bf 70} (2004) 083004.

\end{thebibliography}
\end{document}